# Networks of Josephson Junctions and Their Synchronization

Yurii N. Ovchinnikov[1,2] and Vladimir Z. Kresin[3]

[1]L. Landau Institute of Theoretical Physics, 117334, Moscow, Russia

[2]Max-Planck Institute for Physics of Complex Systems, Dresden, D-01187, Germany

[3]Lawrence Berkeley Laboratory, University of California at Berkeley, CA 94720, USA

**Abstract**

One can demonstrate that a 1-D Josephson network containing junctions with different resistances can be synchronized at frequencies, which are multiples of $2eV$, where $V$ is the total d.c. voltage applied across the network. The appearance of such synchronization follows from the law of charge conservation and takes place if charge transfer is dominated by the Josephson channel. One can observe also a sub-harmonic structure. The result holds for cluster-based arrays as well as for general case of a tunneling network.

## 1. Introduction.

This paper is concerned with Josephson tunneling networks. It represents a continuation of our analysis [1,2] of the networks formed by superconducting nanoparticles. The papers [1,2] describe a transfer of the d.c. current, while here we study the impact of an external voltage applied to the network. More specifically, the focus of the present paper is on the possibility of network synchronization. It should be emphasized that the problem of synchronization is not specific to nanosystems but is of general importance and has attracted a lot of interest (see, e.g., Refs. [3-7])

As is known a single junction radiates at the frequency $\omega = 2eV$, $V$ is the externally applied voltage. If the voltage has d.c. and a.c. components:

$$U = V + v \sin \Omega t \qquad (1)$$

then at certain frequencies, $\Omega = 2eV/n$, Shapiro steps can be observed [8].

Consider 1-D Josephson tunneling network consisting of $N$ junctions and having an external d.c. voltage $V$ applied at the ends of the network. If the junctions are all equivalent, then the network radiates with the frequency $\omega = 2eV/N$.

In this paper we focus on the more realistic case when the junctions are not equivalent. Then one might expect that because of non-uniform distribution voltages across the network a rather broad spectrum of radiated frequencies would be observed. However, the picture is more complicated and, as a whole, this is a complex, non-linear problem. Nevertheless, it can be demonstrated, and this is the main goal of the present paper, that despite the inequality of the junction tunneling resistances, the network can be synchronized at the frequency $\omega = 2eV$, where $V$ is the total voltage. This follows from the law of charge conservation. The microscopic derivation of this statement will be presented below.

**2. Main Equations**

Consider the tunneling 1D network, which contains $N$ junctions. The total voltage $V$ is applied at the ends of the network, and, therefore,

$$V = \sum_{n=1}^{N} V_n \tag{2}$$

Here $V_n$ is the voltage across the $n$-th junction.

As was noted above, $V = const$. As for the voltages $V_n$ ($n = 1, \dots, N$), they, generally speaking, depend on time, so that $V_n \equiv V_n(t)$.

The current $j_n$ flowing through the $n_{th}$ junction has a form

$$j_n = j_n^0 \sin\left(\int_0^t V_n dt + \varphi_n\right) + (V_n/R_n) \tag{3}$$

$$j_n^0 \equiv j_{n;max}$$

The first term describes the Josephson current. The second term corresponds to the one particle current, which is the tunneling of the thermal excitations ($\sim \exp(-\varepsilon_n/T)$). This term is small at temperatures $T \ll \varepsilon_n$, that is, at low temperatures. One can incorporate also the displacement current (see below).

Because of the charge conservation law, the currents flowing through neighboring junctions should be equal, that is $j_s = j_{s+1}$, or:

$$\frac{V_s}{R_s} + J_s^0 \sin\left(\int_0^t V_s dt + \varphi_s\right) = \frac{V_{s+1}}{R_{s+1}} + J_{s+1}^0 \sin\left(\int_0^t V_{s+1} dt + \varphi_{s+1}\right)$$

where $s = 1, \ldots, N - 1$ (4)

Eqs.(2,4) are the main equations, which form the basis for the analysis. As was noted above, we are dealing with a complex, non-linear problem. and our main goal is to demonstrate the possibility for the network to be synchronized. According to Eq.(3), the tunneling current has two components: 1) Josephson tunneling and 2) one-particle ("normal") component. Especially interesting is the situation when the Josephson channel dominates the charge transfer ($T \ll \varepsilon_n$).

### 3. Synchronization

Consider a Josephson tunneling network with an external d.c. voltage *V* applied at the ends. We focus on the most interesting case when the Josephson channel dominates the charge transfer across the network. As mentioned above, we consider the realistic situation of inequivalent Josephson junctions. Let us demonstrate that such a network can be synchronized at frequencies determined by the external potential *V*. One also can show (see below) that the displacement current as well as the additional contribution from the one-particle tunneling does not affect this synchronization.

At first, let us analyze the network containing just two junctions. This case is interesting for its own sake and, in addition, allows us to demonstrate the main physics of the synchronization phenomenon.

Then we have (See Eqs.(2),(4))

$$V = V_1 + V_2 \qquad (5)$$

$$J_1^0 \sin\left(\int_0^t V_1 dt + \varphi_1\right) = J_2^0 \sin\left(\int_0^t V_2 dt + \varphi_2\right) \qquad (6)$$

We consider the case of two different junctions. Assume that $J_1^0 > J_2^0$. With use of Eq. (5), Eq. (6) can be written in the form:

$$\eta \sin Z_1 = \sin(2eVt - Z_1 + \theta) \qquad (7)$$

here

$$Z_1 \equiv Z_1(t) = \int_0^t V_1 dt + \varphi_1; \; \theta = \sum_{i=1}^2 \varphi_i; \; \eta = J_1^0/J_2^0 > 1$$

As a result, we obtain:

$$Z_1 \equiv 2e \int_0^t V_1 dt + \varphi_1 = arctg \frac{\sin[2eV(t+t_0)]}{\eta + \cos[2eV(t+t_0)]} \qquad (8)$$

where $t_0 = \theta/2eV$.

One can see directly from Eq.(8) that the function $Z_1(t)$ and, correspondingly, the current $j_1$ (see Eqs.(3),(6)) is the periodic function with frequency $\omega = 2eV$. Because of it, the quantity $Z_1$ defined by Eq.(8) can be expanded in a series

$$Z_1 \equiv 2e \int_0^t V_1 dt + \varphi_1 = \sum_{n=1}^\infty \frac{B_n}{n\omega} \sin n\omega(t + t_0) \qquad (9)$$

where $\omega = 2eV$, and

$$\frac{B_n}{n\omega} = \frac{1}{\pi} \int_0^{2\pi} dz \arctan[\sin z(\eta + \cos z)^{-1}] \sin nz \qquad (9')$$

One can see also from Eq.(9) that

$$2eV_1(t) = \sum_{n=1}^\infty B_n \cos n\omega(t + t_0) \; ; \; \omega = 2eV \qquad (10)$$

Therefore, indeed, the potential $V_1$, as well as $V_2 = V - V_1$, depends on time.

One can see also that

$$Z_2 \equiv \int_0^t V_2 dt + \varphi_2 = 2eV(t+t_0) - Z_1 \qquad (11)$$

$Z_1$ is determined by Eq.(8) and is described by the expansion (9) with coefficients (9').

Therefore, the network consisting of two different junctions ($J_1^0 \neq J_2^0$) is synchronized with frequencies $\omega = 2eV, 4eV, ...$ which are multiples of the external total potential.

As was noted above, the potential $V_1$, and $V_2$ depend on time, so that $V_1 \equiv V_1(t)$ and $V_2 \equiv V_2(t)$.

It is essential that the average values of the potentials $V_1$ and $V_2$ are different. Indeed, one can see directly from the definitions (8), (9) and the expression (5), that $<V_1> = 0$, whereas $<V_2> = V$. Therefore, the charge conservation law (Eq.(6)) leads to rather peculiar time dependent distribution of an external potential.

If the junctions are very different ($\eta \gg 1$), then, as follows from Eq.(9'): $B_1/\omega \simeq \eta^{-1}$, for $n=1$, and $B_n \ll B_1$ for $n \neq 1$ (e.g., $B_2/\omega \approx -\eta^{-2}$, etc.).

Based on Eqs. (8),(9), one can see that if the external field has a general form (1), it leads to an appearance of the Shapiro steps. However, the picture of the Shapiro steps is more complex relative to that for the single junction. Namely, in addition to main steps at $\hbar\Omega = n\hbar\omega$ ($\hbar\omega = 2eV$), one can observe the subharmonic structure. Indeed, for a single junction the current is described by the simple sinusoidal dependence. If we are dealing with the array, the dependence is also periodic, but is more complicated (see Eqs. (8), (9)). In this case one should perform an additional Fourier transformation (Eqs. (9), (10)), and we obtain the picture of the Shapiro steps, so that $2eV = \hbar\Omega n/n'$. The scenario is similar to that for superconducting microbridges (see e.g. [9] and also review [10]), when the current is also described by the periodic function different from the simple dependence. The I-V characteristics will display usual Shapiro steps at n=1,2,…, but, in addition, one can observe sub-harmonic structure corresponding to various values of n'.

In a similar way one can consider the network , which contains 3 junctions. In this case we are dealing with the equations:

$$\eta_{1;2} \sin Z_1 = \sin Z_2 \qquad (12)$$

$$\eta_{2;3} \sin Z_2 = \sin Z_3 \qquad (12')$$

$$V = \sum_{i=1}^{3} V_i \qquad (12'')$$

where $Z_i = 2e \int_0^t V_i dt + \varphi_i$ ; $i = 1,2,3$, $\eta_{r;s} = J_r^0/J_s^0$; $r; s \equiv 1,2,3$ (cf. Eqs.(4)); Assume that $j_1^0 > j_2^0 > j_3^0$. One can see that

$$Z_3 = 2eV(t + t_0) - Z_1 - Z_2; \quad t_0 = \sum_{i=1}^{3} \varphi_i/2eV.$$

With use of Eqs. (12'), (12''), we obtain

$$Z_2 = arctg \frac{\sin[2eV(t+t_0)-Z_1]}{\eta_{2;3}+\cos[2eV(t+t_0)-Z_1]}$$

(13)

Using this expression and Eq. (12), one can express the potential $V_1$ in terms of $V$:

$$\sin Z_1 = \eta_{3;1} \frac{\sin[2eV(t+t_0)-Z_1]}{\{1+\eta_{3;2}^2+2\eta_{3;2}\cos[2eV(t+t_0)-Z_1]\}^{1/2}} \qquad (14)$$

Therefore, all quantities $Z_1, Z_2, Z_3$ depend periodically on time with the frequency ω= $2eV$.

Once again, one can see that the system can be synchronized with an external potential $V$ applied on the network's edges. The main

Shapiro steps again will be multiples of $V$ ($\Omega = 2eVn, n = 1,2,...$); there will be also a sub-harmonic structure, (cf. [10]), see above. If $\eta_{3;2} \ll 1$, we obtain from the Eqs.(12):

$$Z_1 = \eta_{3;1} \sin 2eV(t + t_o) \; ; \; Z_2 = \eta_{3;2} \sin 2eV(t + t_o) \; ;$$
$$Z_3 = 2eV(t + t_o) \tag{15}$$

Therefore, indeed, this network is synchronized with an external voltage $V$. Note also that $< V_1 >=< V_2 >= 0$, whereas $< V_3 >= V$.

The derivations can easily be generalized for a general case ($N$ junctions). One can analyze the system

$$J_s^0 \sin\left(\int_0^t V_s dt + \varphi_s\right) = J_{s+1}^0 \sin\left(\int_0^t V_{s+1} dt + \varphi_{s+1}\right)$$

$$(s = 1, ..., N - 1)$$

which is valid for the case when Josephson channel is dominant. By analogy with the derivation described above, one can demonstrate that all currents are periodic functions of the total external potential $V$; this leads to corresponding synchronization and an appearance of the Shapiro steps.

Note that if we consider the small additional contribution of the one-particle tunneling (see Eqs (3),(4)), it would not affect the synchronization picture. Indeed, consider again the system containing two junctions. Based on Eq. (4), one can write (cf. Eq. (6))

$$\eta \sin Z_1 = \sin[2eV(t + t_0) - Z_1] + S \tag{16}$$

where $Z_1, t_0, \eta$ are defined by Eqs. (7), (8), and

$$S = [(V - V_1)R_2^{-1} - V_1 R_1^{-1}]/J_2.$$

Since $S \ll 1$ in the low temperature region ($T \ll \varepsilon_i$), one can treat this term as a perturbation. Correspondingly, one can write

$V_1 = V_1^0 + V_1^1$, so that $V_1^0$ is described by Eqs. (9)-(10). After simple calculation, we obtain:

$$Z_1^1 \equiv 2e \int_0^t V_1 dt + \delta\varphi_1 = \tilde{S}\{\eta \cos Z_1^0 + \cos[2eV(t + t_0) - Z_1^0]\}^{-1} \tag{17}$$

where $f^0 = \cos[2eV(t + t_0) - Z_1^0]$, $\tilde{S} = S + \delta\theta f^0$

It is essential that the denominator in the right side of Eq. (17) is not equal to zero at any value of $V$ and $t$. Indeed, it could become equal to zero if $tg Z_1^0 = -[\eta + \cos 2eV(t + t_0)]/\sin 2eV(t + t_0)$, which is incompatible with the solution (8). Therefore $Z_1^1 \ll 1$ and is a periodic function of the external potential $V$; it follows from (9), (14).

As a result, the system will be synchronized. This conclusion is valid also if we take into account the displacement currents $C_i(\partial V/\partial t)$. We assume that its contribution is also small relative to that of the Josephson channel. In other words, we assume that the parameter $(CV\omega/J_c^0) \ll 1$; such a case is perfectly realistic (see below). The treatment can be performed similarly to that for the one-particle tunneling (see above). One can see that in this case the correction to $Z_{1;dc}$ (cf. Eq. (15)) has a form:

$$Z_{1;dc} = (C_1 + C_2)J_2^{-1}[\eta + \cos(2eV(t+t_0))\cos Z_1 + \sin(2eV(t+t_0))\sin Z_1]^{-1} \partial V_1^0/\partial t$$

(18)

and is periodic with $\omega = 2eV$.

## 4. Discussion

It has been demonstrated above that 1-D network consisting of Josephson junctions can be synchronized at a frequency equal to $\omega = 2eV$, where $V$ is the external voltage applied across the network. Correspondingly, one can observe the Shapiro steps. Thus, the network response is similar to that of a single junction. The main condition, which needs to be satisfied to attain the described synchronization is that the Josephson tunneling channel must be dominant. That is, one-particle tunneling must be relatively small ($T \ll \varepsilon_i$, where $\varepsilon_i$ is the energy gap).

The displacement current should be also relatively small. Such a situation is perfectly realistic. For example, if the network is characterized by the following values of the parameters: $V \simeq 2.2mV$, $C \approx 0.5pF$, $\omega \approx 5 \times 10^2 GHz$, $J_c^0 \approx 3 - 5mA$, then $(CV\omega/J_c^0) \ll 1$ ($\sim 10^{-1}$).

Network synchronization has been observed in a number of studies (see, e.g., [5-7]). Measurements were performed, mainly, with 1-D and 2-D arrays where the junctions were similar or almost similar. For the 1-D network of N such junctions the synchronization occurs at $\omega = 2eV/N$. Here we focus on the entirely different case of junctions that are far from similar. It is remarkable that in this a case one still can observe synchronization (and Shapiro steps) at frequencies, which are multiples of $2eV$, where $V$ is the total d.c. voltage across the network. The analytical treatment described above shows that this phenomenon occurs thanks to a peculiar voltage distribution, which is time dependent and follows directly from the principle of charge conservation.

The synchronization of couple YBa$_2$Cu$_3$O$_{7-\delta}$ junctions was described in [11],[12]. The Shapiro steps were observed when 12GHz microwave radiation with sufficient power was applied. The I-V characteristics clearly display the Shapiro steps if $V = \hbar\Omega/2en$, where $V$ is the voltage across both junctions. We think that these steps correspond to the synchronization described above.

Let us also make a comment on the "giant" Josephson proximity effect observed in [13]. The electrodes were formed by La$_{0.85}$Sr$_{0.15}$CuO$_4$ films ($T_c \simeq 45K$) and were separated by the underdoped LaCuO compound with $T_c' \simeq 25K$; the width of the separating layer was 200Å (!). Despite such a large scale of separation the authors of [13] observed $at\ 35K > T > T_c'$ the d.c. and a.c. Josephson tunneling. According to our theory [14, 15], the separating layer contains superconducting regions embedded in a normal metallic matrix, hence we are dealing with Josephson tunneling through a network of such regions. Our analysis of the d.c. Josephson current [16] is in a good agreement with the data [13]. According to [13], one can observe the Shapiro steps, so that the picture appears to be similar to that for a single junction. We think that the observation reflects the synchronization described above. The superconducting state in the isolated regions persists up to $T_c^* \simeq 80K$ [17], so that the condition $T < \varepsilon$ is satisfied.

As was noted in the Introduction, the present paper is a continuation of our previous work [1], [2]. The present study permits one to state that a cluster-based Josephson network is also capable of transferring an a.c. current synchronized at the frequency corresponding to the total voltage applied across the network.

The authors are grateful to R. Dynes and S. Cybart for fruitful discussions. The research of YNO is supported by EOARD, Contract No. 097006.